\documentclass[fleqn,usenatbib]{mnras}

\usepackage{newtxtext,newtxmath}

\usepackage[T1]{fontenc}
\usepackage{ae,aecompl}

\usepackage{graphicx}	
\usepackage{amssymb}	

\title{Examining supernova events in Type 1 active galactic nuclei}

\author[Villarroel et al.]{
Beatriz Villarroel,$^{1,2}$\thanks{E-mail: beatriz.villarroel@su.se}
I\~nigo Imaz,$^{3}$
Elisabeta Lusso$^{4,5}$, S\'{e}bastien Comer\'{o}n$^{6,2,8}$ \and M. Almudena Prieto$^{2,8}$, Paola Marziani$^{7}$, Lars Mattsson$^{1}$
\\
$^{1}$Nordita, KTH Royal Institute of Technology and Stockholm University, Roslagstullsbacken 23, SE-106 91 Stockholm, Sweden\\
$^{2}$Instituto de Astrofisica de Canarias, Avda Via Lactea S/N, La Laguna, E-38205, Tenerife, Spain\\
$^{3}$Department of Physics and Astronomy, Uppsala University, Uppsala, Sweden\\
$^{4}$Dipartimento di Fisica e Astronomia, Universit\`a di Firenze, via G. Sansone 1, 50019 Sesto Fiorentino, Firenze, Italy \\
$^{5}$INAF--Osservatorio Astrofisico di Arcetri, 50125 Firenze, Italy\\
$^{6}$Space Science and Astronomy, University of Oulu, P.O. Box 3000, FI-90014 Oulu, Finland\\
$^{7}$INAF, Osservatorio Astronomico di Padova, Italy\\
$^{8}$Departamento de Astrofísica, Universidad de La Laguna, E-38205, La Laguna, Tenerife, Spain
}

\date{Accepted XXX. Received YYY; in original form ZZZ}

\pubyear{2019}

\begin{document}
\label{firstpage}
\pagerange{\pageref{firstpage}--\pageref{lastpage}}
\maketitle

\begin{abstract}
A statistical study of intermediate Palomar Transient Factory supernovae (SNe) in Type 1 AGN has shown a major deficit of supernovae around Type 1 AGN host galaxies, with respect to Type 2 AGN hosts. The aim of this work is to test whether there is any preference for Type 1 AGN to host SN of a specific kind. Through the analysis of SN occurrence and their type (thermonuclear vs core-collapse), we can directly link the type of stars producing the SN events, thus this is an indirect way to study host galaxies in Type 1 AGN. We examine the detection fractions of SNe, the host galaxies and compare the sample properties to typical host galaxies in the Open Supernova Catalog \citep[OSC; ][]{Guillochon2017}. The majority of the host galaxies in the AGN sample are late-type, similar to typical galaxies hosting SN within the OSC. The findings are supportive of a deficiency of SNe near Type 1 AGN, although we cannot with certainty assess the overall detection fractions of SNe in Type 1 AGN relative to other SN host galaxies. We can state that Type 1 AGN has equal detection fractions of thermonuclear vs core-collapse SNe. However, we note the possibility of a higher detection rate of core-collapse supernovae in Type-1 AGN with insecure AGN classifications.

\end{abstract}




\section{Introduction}

An Active Galactic Nucleus (AGN) is an extremely luminous galactic core, driven by accretion of gas upon a super-massive black hole \citep{LyndenBell1969,Rees1984}. AGN are thoroughfully studied as they are believed to play a major role in the evolution of galaxies through so-called 'AGN feedback', where star formation in a galaxy may be quenched or even stimulated by the presence of an AGN, see e.g. \cite{King2015}. Studying the host galaxies of AGN is thus key to both learn about the intrinsic physics of compact objects and how these powerful cores influence the environment they are in (the galaxy itself) and vice versa. Moreover, AGN display a large variation in luminosity and numbers over redshift (with a peak in both number and emission at $z \sim 2$), while they have almost entirely vanished from our Local Universe ($z \sim 0$). Therefore, their demographics give us important clues about the structure formation in the Universe and black-hole/galaxy co-evolution.

In the optical, AGN are usually split into two major groups; those having spectra with Doppler-broadened emission lines as the observer has an unobscured view directly into the accretion disk that is surrounded by fast-moving gas clouds ('Type 1 AGN'), and those having narrow Balmer lines as dust blocks the line-of-sight into the accretion disk \citep['Type 2 AGN';][]{Antonucci1993}. This dust obscuration comes from dust at different scales, starting from the smallest regions near the accretion disk, and extending to dense dust in the host galaxies.
For instance, Type 2 hosts are associated with higher star-formation \citep{Maiolino1995,Koulouridis2013,Villarroel2017} and more dust obscuration \citep{Malkan1998} than Type 1s. Moreover, the increased ratio of Type 1 /  Type 2 AGN at higher AGN luminosities \citep{Lusso2013} and different 'engine power' \citep[e.g.][]{Villarroel2017,Ricci2017} between the two AGN groups suggest they may be probing different stages in the lifetime of an AGN.

In this study, we construct a sample of SNe exploding in the host galaxies of optically-selected Type 1 AGN. SNe provide the least contaminated window into the star-formation history of the host galaxy. Depending on the detection fractions and classes of SNe, we may have hints on the stellar masses or the average ages of the stars. Core-collapse SNe are typically linked to massive young stars, while thermonuclear SNe are directly connected to an older stellar population. Indeed, host galaxies of AGN show less star-formation on average than non-active galaxies \citep{Nazaryan2013,Hakobyan2014} but it is connected to the morphologies of the AGN hosts being earlier type. In a previous work \citep{Villarroel2017}, we used SNe from the coherently collected intermediate Palomar Transient Factory (iPTF) to study the SNe in Type 1 and Type 2 AGN host galaxies. We found that AGN hosts on average had lower star-formation rate than normal star-forming galaxies, but more interestingly, only one single SN was found among all Type 1 AGN hosts.

Another interesting use-case of such a Type 1 AGN / SN sample, is that one can directly connect two independent ways of measuring the cosmological distance. While thermonuclear (Type Ia) supernovae have been used as standard candles up to redshifts $z <$ 1.8, only recently a method
to use quasars as a standard candle has been developed and refined with larger samples of quasars \citep{Risaliti2015,Risaliti2017}. The method utilises a non-linear relationship between the X-ray and UV luminosities of quasars at 2 keV and 2500 \AA, and already has been used for cosmological
tests (e.g. \cite{LopezCorredoira}). In a recent work \citep{Risaliti2018}, the quasar Hubble diagram showed an excellent agreement with the Hubble diagram based on TypeIa supernovae and the concordance model. More interestingly, at even higher redshifts where supernovae no longer are observed, the quasar Hubble diagram reveals an interesting deviation from the CDM, that may suggest that the dark energy density increases at higher redshift. Therefore, constructing a sample of Type 1 AGN with past supernova events is valuable.

In this paper, we construct a sample of all documented SN cases in the scientific databases that have been reported in Type 1 AGN to study their host galaxy properties and the general properties of our sample. The constructed sample, may be particularly useful for future studies aiming at developing cosmological distance indicators, as the thermonuclear supernovae often have associated distance moduli.

\section{Sample}\label{sec:Sample}
\begin{figure*}
	\includegraphics[scale=0.55]{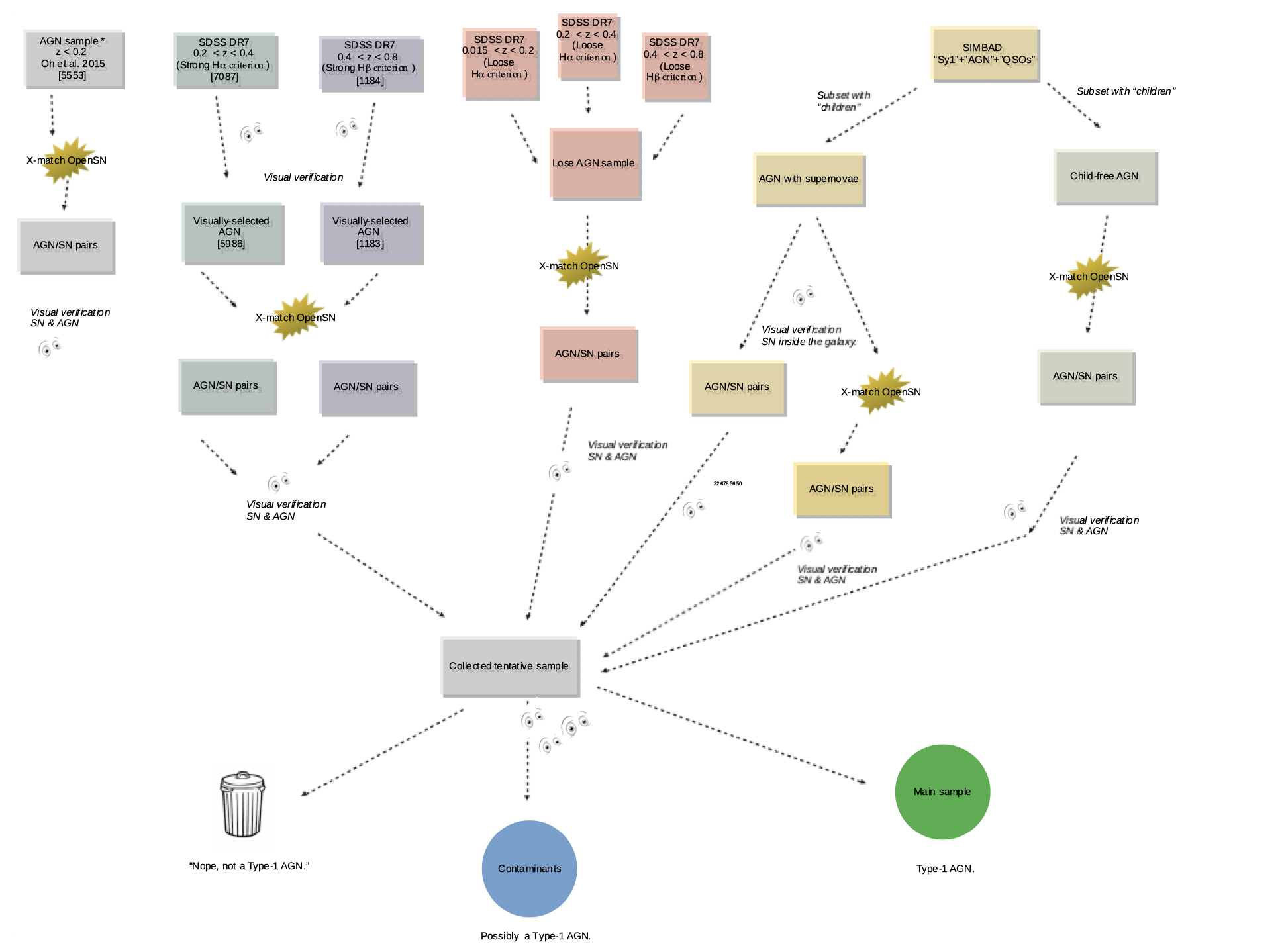}
    \caption{Flowchart demonstrating the overall sample selection, as described in Section \ref{sec:Sample}.}
    \label{Flowchart}
\end{figure*}

Here we present the approach we adopted to search for SNe hosted in Type 1 AGN from various archives. The overall scheme can be seen in Figure \ref{Flowchart}.

\subsection{Initial SN selection}

As we are interested in finding the largest possible number of AGN-SN pairs to build our sample, we use the OSC, the version of 3/08/2018. The OSC is collected from a wide range of sources and online catalogs and simply represents ``all what can be found'', currently collecting about 55 000 recorded events. For a AGN/SN table with a purpose to be useful in e.g. tests of various cosmological distance indicators, the OSC is well suitable.

However, for sample comparison purposes when one chooses to compare SNe frequency in different classes of objects, one must remember this catalogue has a incoherent (biased) collection of the SNe that prioritizes the detection rate over completeness with respect to SN types or galaxy types. On the other hand, if one uses the OSC itself as a control sample, it is possible to compare it to a single class of objects.

We have selected all supernova at $z < 0.8$, removing the unconfirmed SN candidates (that include contaminants like minor planets or luminous blue variables) one gets a sample of 34893 SNe. Of these 34893  SNe, 15387 have known redshifts, and the remaining 19506 do not have a known redshift.

\subsection{The SDSS samples}\label{ref:initialAGN}

To construct a sample of Type 1 AGN that all have shown a SN going off, we must resort to several different approaches, using both the Sloan Digital Sky Survey (SDSS) \citep{York2000} Data Release 7 (DR7) \citep{Abazajian2009} and the SIMBAD archive. The SDSS DR7 was chosen as the catalogue provides us with the spectral fitting properties (e.g. FWHM, continuum line flux, line widths) for all the objects. With help of the tables 'SpecLine' and 'SpecObj', we can use open criteria without restricting ourselves to any SDSS subcatalogues.

For the Type 1 AGN, we use different sample selection methods to select the Type 1 AGN.

\begin{enumerate}
\item We directly used the final Type 1 AGN sample from \cite[Sec. 2.3;][]{Oh2015}, that has 5553 Type 1 AGN in the redshift range 0 $< z < $ 0.2.\footnote{We thank K. Oh for generously providing us the sample.} The Type 1 AGN in \cite{Oh2015} were required to have an amplitude-over-noise ratio larger than 3, full width half maxima in H$\alpha$ larger than 800 km/s and were identified using the OSSY catalogue that has carefully remeasured emission lines \citep{Oh2011}. The OSSY catalogue uses line measurement methods from \cite{Sarzi2006}. The extensive way of selecting Type-1 AGN in \cite{Oh2011} allows many Type 1 AGN that normally go undetected due to inaccuracies in line measurements to be identified.

\item 'Strong criteria': Using the CasJobs Interface and SDSS DR7 ('SpecLine') we select extragalactic objects with $\sigma$(H$\alpha$) $> 15  $ \AA\ and within 0.015 $< z <$ 0.4. Then for objects with $\sigma$(H$\beta$) $> 15$ \AA, and within 0.4 $< z <$ 0.8. The first search resulted in 7087 Type 1 AGN, and the higher-redshift search in 1184 Type 1 AGN. Visually examining the AGN  and removing false positives with inaccurately reported line widths, leaves 5986 objects within 0.2 $< z <$ 0.4, and 1183 within 0.4 $< z <$ 0.8.

\item 'Loose criteria': we select objects within 0.015 $< z <$ 0.2 and 0.2 $< z <$ 0.4 with $\sigma$(H$\alpha$) $> 10  $ \AA\, resulting in 41125 and 20724 Type-1 AGN. We also select objects $\sigma$(H$\beta$) $> 10  $ \AA\ from 0.4 $< z <$ 0.8, giving us 17572 objects. Here, we do no visual classification before the cross-matching with the OSC.

\end{enumerate}

The samples above, we refer to as the 'SDSS parent samples', are then cross-matched as described in Section \ref{sec:crossSDSS}.

For parent samples with ``Strong'' and ``Loose'' criteria, we chose to restrict ourselves in the redshift regime of 0.015 $< z < $ 0.8. The higher limit we chose as we require a precise redshift from the narrow [OIII] emission line, but also due to the deficiency of SN detections at high redshifts ($\sim$ 75\% of all detected SNe are found at low $z < 0.2$).

We make a cross-match of the AGN in the ``Loose parent sample'' with $z < 0.2$ using the CDS Xmatch Service and a 2 arcsec cross-match radius and find that 23838 out of 41125 Type 1 AGN can be recovered in the SIMBAD catalogue, meaning that the overlap is about $\sim$ 60\%.

\subsubsection{Cross-matching SNe with the SDSS parent samples}\label{sec:crossSDSS}

For the SDSS parent samples we set two necessary conditions to consider a SN having a known redshift to be associated with a given AGN.

\begin{enumerate}
    \item A redshift condition $\left|z_{AGN} - z_{SN}\right| < 0.005$. Even with a less strict condition of $\left|z_{AGN} - z_{SN}\right| < 0.01$
    one can safely say that 97\% of the objects within each pair are associated with each other \citep{Wang2010}.
    
    \item Each SN must be found within a search radius projected distance less than 100 kpc, and an angular distance less than 3 arc minutes. The projected distance is calculated using a Hubble constant of $h_{0}$ = 72 km s$^{-1}$ Mpc$^{-1}$.

\end{enumerate}

Using a similar approach in \cite{Villarroel2017}, we showed that most objects in each pair were associated with each other, as could be seen by a bottom-heavy distribution of a histogram showing the number of pairs with different $\left|z_{AGN} - z_{SN}\right|$. For the SNe that have unknown redshifts, we set the assumption that the redshift of the SN is the same as of the galaxy.

For example, from the parent samples with 'Loose criteria', we initially find 150 potential Type 1 AGN/SN pairs. As some of the SNe might go off in a nearby neighbour, we visually inspect the images in the SDSS Explorer and only keep the AGN-SN pairs were the SN has gone off inside the host galaxy of the AGN.

\subsection{The SIMBAD samples}

In order to extend our sample of AGN-SN pairs and include objects also from the southern hemisphere, we have also used the SIMBAD archive to select the Type 1 AGN, see Section \ref{sec:visual}.

The Type 1 AGN can be found under the SIMBAD categories ``AGN'', ``Sy1'' or ``QSO'', each of these categories containing tens of thousands of objects (``Sy1'': 20.000+, ``AGN'': 39.000+, ``QSO'':307.000+). As the SIMBAD is a digital archive, many of these objects may be misclassified as Type 1 AGN due to the automatized classification, which will require a visual inspection at later stages.

\subsubsection{Cross-matching with the SIMBAD parent samples}\label{sec:crossSIMBAD}

To efficiently cross-match these samples in searches for AGN/SN pairs, one might either do similarly as in the Section \ref{sec:crossSDSS}), or take advantage of the ``children'' function in SIMBAD query system. When a galaxy has something detected inside of it, it should (in principle) have ``children''. By filtering on the AGN that have ``children'', we may directly find the most probable candidates to have SNe associated. This gives us initially 89 `QSOs', 192 `AGN' and 119 `Sy1' with ``children''.

After getting a first selection of AGN with ``children'', we remove the hits that contain non-SN children and calculate the distances between every potential AGN and SN, and visually examine if the possible SN is localised within the AGN host galaxy.

After this, we also take all the Type 1 AGN without ``children'', and cross-match with the OSC and visually inspect. We do this to ensure that we do not lose potential cross-matches by relying on the children function. As some of the host galaxies with ``children'' may have several SN, we also take all the Type 1 AGN that had ``children'' listed and cross-match then with the OSC.
This numbers here are 158 `QSOs', 692 `AGN' and 159 `Sy1' that initially appear to have a nearby SNe.

Since there is still a possibility that we lose some AGN-SN matches by this approach, finally, we cross-match the complete AGN samples from SIMBAD with the OSC, and again only keep those that have a SN located within the host galaxy. We store the host galaxy names and coordinates. Figure \ref{Flowchart} summarizes all the ways of cross-matching our samples to ensure that we find all listed objects.

The goal of the next step, the visual verification, is to remove all the false pairs.

\subsection{Visual verification of the candidate pairs}\label{sec:visual}

Even if we have a list of possible AGN-SN matches, we do know that most of these will be contaminants of some sort. In particular for the Type-1 AGN originating from the 'loose' samples, we can anticipate many of these AGN to be normal galaxies or Type 2 AGN with inaccuracies in the line width measurements given by the SDSS DR7 pipeline. An example is a high-redshift galaxy where the Lyman alpha has been confused with Balmer lines in DR7, and that sneaked into the sample mistakenly.

\begin{figure*}
   \centering
  \includegraphics[scale=0.3]{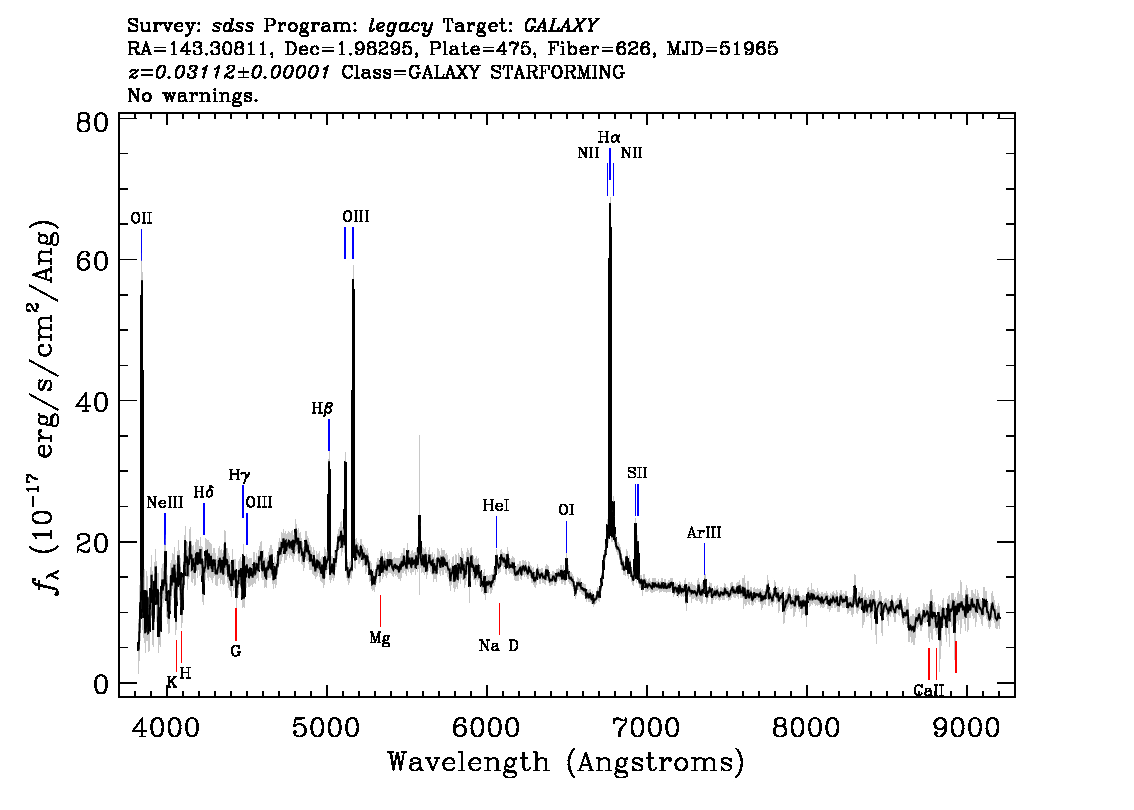}
  \includegraphics[scale=0.3]{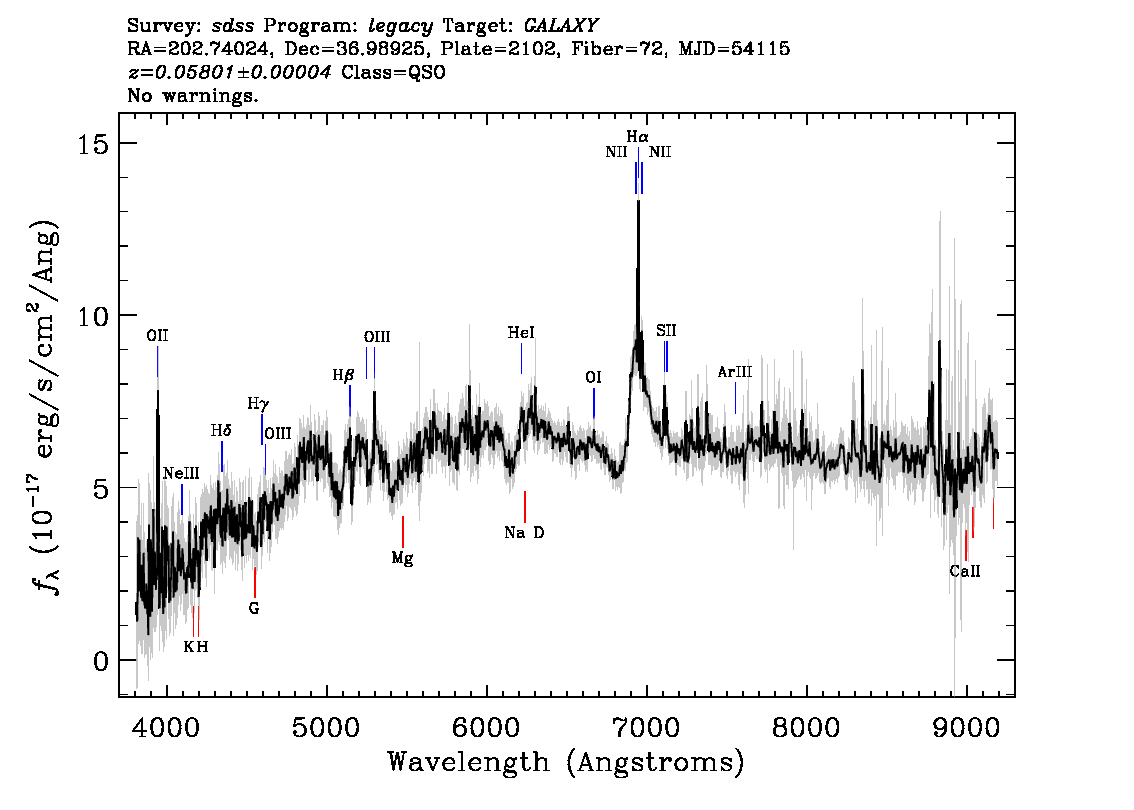}
  \caption{\label{voigt} {Examples of dubious Type-1 AGN candidates. The upper image shows the spectrum of the host galaxy (VV2010c) J093313.9+015858 and the lower image shows the spectrum of the host galaxy NGP9 F270-0377962. Both candidates passed the initial candidate selection criteria based on Balmer line width, but may be normal galaxies with superposed SNII spectra rather than any real Type-1 AGN.}}
   \end{figure*}

For each of our Type 1 AGN-SN pairs, we therefore must examine the AGN spectra in the SDSS and the NASA/IPAC Extragalactic Database (NED) databases, in order to remove all contaminants. Among the contaminants are objects that might be similar to or higher than Type-1.5 objects (e.g. 1.8). Some ordinary galaxies with other supernova activity going on may also get broadened line profiles, confusable with Type 1 AGN. We include two possible examples, see Section \ref{comments} further down. Since we are not sure, we consider also them possible contaminants.

Also, some of our reported SNe might be false alarms. With the help of literature, we do a check and remove those that may have been misreported as SNe in the OSC. As an example, we remove SN 1984Z in NGC 5548 that is an unverified transient and move it to a separate sample. Since Mrk 590 is a changing-look AGN that switches between Type 1 and Type 2 states, we move the transient 2018djd into this separate sample with possible contaminants. We also move the unverified transients 1999fq and 1990ad that reside in two less well known galaxies into this separate sample. The other AGN/SN pairs we are unsure about, we store in this separate table that we will compare with our main table. We use the contaminant table later to compare the results with and without contaminants. The main and contaminant tables are found in Tables \ref{SupernovaCollection} and \ref{Contaminants}.

More details on candidate selection can be found in the master thesis of \cite{Imaz2019}.

\subsubsection{Two dubious Type-1 AGN}\label{comments}

Two examples of objects that may have broad lines but be impostors, are the galaxies (VV2010c) J093313.9+015858 and NGP9 F270-0377962. Their spectra are shown in Figure \ref{voigt}.

The P-Cygni profiles observed in the H$\alpha$ profiles of (VV2010c) J093313.9+015858 and NGP9 F270-0377962 are almost never associated with quasar spectra. An extremely rare class of quasars shows broad absorptions in the Balmer lines \citep{Hall2007}, with a wide range of absorption through \citep{Zhang2015}. Quasars with broad Balmer absorption lines are luminous, and often classified as low-ionization BAL QSOs (LoBAL or FeLoBAL) from their UV spectrum \citep{Schulze2018}, which is not the case of  (VV2010c) J093313.9+015858 and NGP9 F270-0377962.  The spectral properties of the two sources appear to be consistent with SN II because of three features: the (a) P-cygni profiles; (b) the absorption in the Na I D band at 5890 \AA; (c) the appearance of the H$\beta$ spectral region between 4500 and 5500 \AA\ which has been seen in SN 1987A at age $\sim$ 5 months \citep[see Fig. II,][]{Filippenko1997} and in SN II supenovae. The spectra can be explained by three fundamental components: (a) the host galaxy, which is clearly detected in the blue part of the spectrum via the HK CaII absorptions whose redshift is in agreement with the narrow emission line spectrum; (b) the absorption/emission spectrum of a SN II; (3) a narrow emission line spectrum. If this interpretation is correct, neither object can be classified as a Type-1 AGN. The evidence of non-stellar nuclear activity has to be based on the narrow emission line spectrum which is relatively easy to measure. We estimated the fluxes and the width of the strongest emission lines used in the diagnostic diagrams of \cite{BPT}, and \cite{viejo}, namely H$\alpha$, H$\beta$, [OIII]$\lambda$5007, [NII]$\lambda$6584, [SII]$\lambda\lambda$6717,6730, and [OII]$\lambda$3727. In the case of VV2010c, diagnostic diagrams based on [OIII]/H$\beta$ vs. [OII]/H$\beta$\ and and [OIII]/H$\beta$ vs. [NII]/H$\alpha$\ positively identify the emitting regions as HII. The only discordant diagnostic diagrams is [OIII]/H$\beta$ vs. [OI]/H$\alpha$\ due to the significant [OI] emission. The extremely low FWHM $\lesssim$ 200 km/s also disfavors the possibility that the emission line spectrum is due to the NLR of an AGN. The case of NGP9 F270-0377962 is different. This source is placed at the boundary between HII and transition objects in the d.d. [OIII]/H$\beta$ vs. [NII]/H$\alpha$; in the [OIII]/H$\beta$ vs. [OI]/H$\alpha$\ and [OIII]/H$\beta$ vs. [SII]/H$\alpha$\ d.d. the source is classified as a LINER. In the diagram [OII]/[OIII] vs. [OIII]/H$\beta$ the data point is not localted along the correlation expected for HII region. The FWHM $\sim$ 400 km/s is also consistent with some low-ionization nuclear activity. In summary, these sources show no evidence of being Type-1 AGN, and only one of them shows some evidence of non-stellar nuclear activity.

\begin{figure*}
	\includegraphics[scale=0.6]{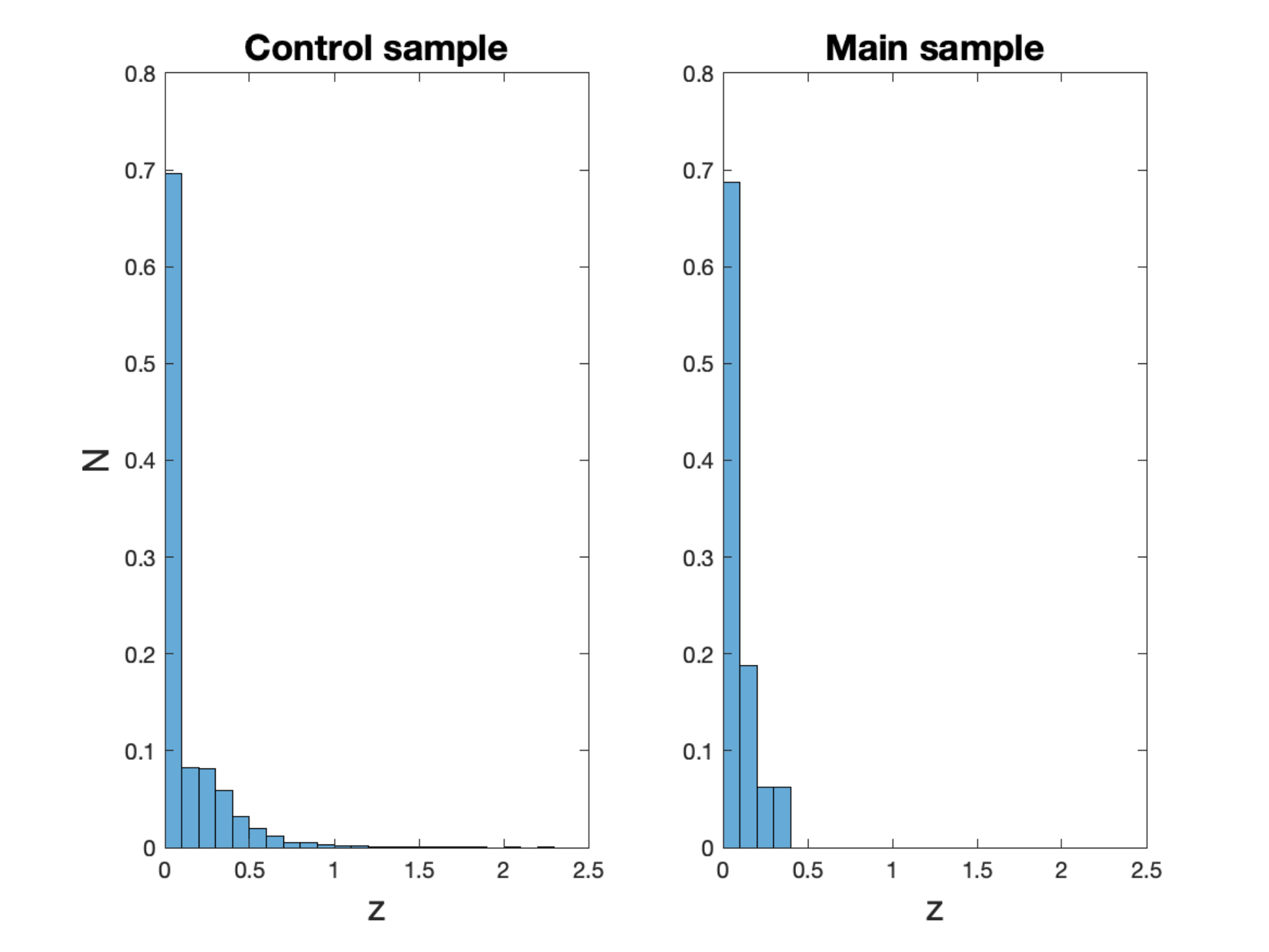}
    \caption{Redshift distributions of the control and main samples. Each plot uses the same bin width and normalisation ('N' is normalised counts). The left plot shows the control sample (10942) and the right plot shows the main sample (16). This plot only includes the events with well-measured redshifts of the SNe.}
    \label{fig:redshift}
\end{figure*}

\begin{figure*}
	\includegraphics[scale=0.6]{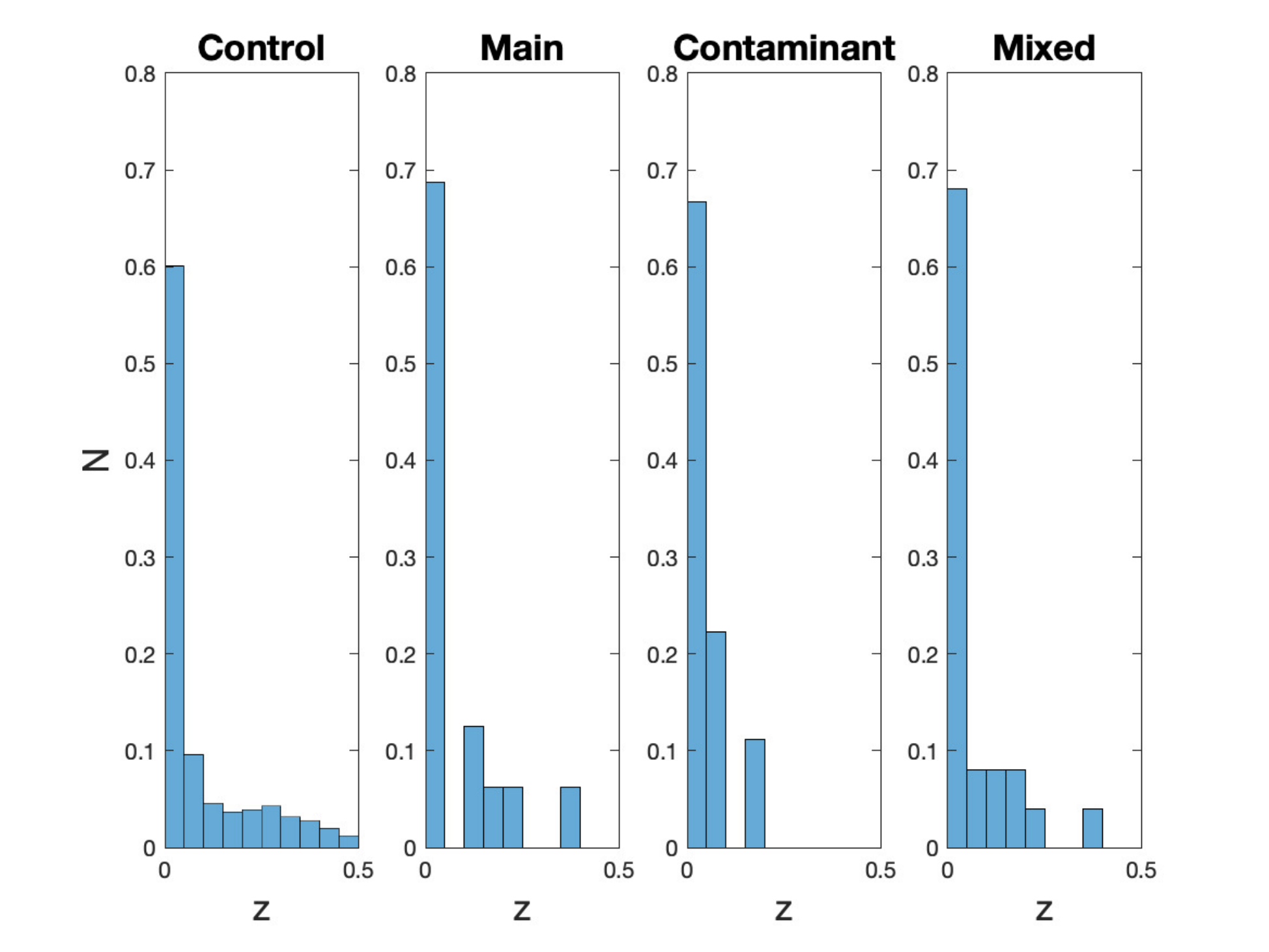}
    \caption{Redshift distributions for pairs at $z < 0.4$. We compare the control sample (10942), the main sample (16), contaminants only (9), and the mixed (25) sample. Only events with well-measured SN redshifts are included. Each plot uses the same bin width and normalisation ('N' is normalised counts).}
    \label{fig:redshiftcont}
\end{figure*}


\subsection{Control sample}\label{sec:control}

To test whether the host galaxies within the sample of Type 1 AGN-SN pairs are peculiar, we selected a control sample of sources within the OSC sample as follows.

We cross-match the confirmed host galaxies from the OSC with the SIMBAD, SDSS and Galaxy Zoo \citep{Lintott2008,Lintott2011} to get the morphologies, radii of the galaxies, position angles and $b/a$ ratios. In total, we obtain this information for 14145 SN host galaxies from the OSC. We list the morphologies for the full sample in Table \ref{Morphologies}. 

The total number of SN events that happen in well identified host galaxies are 16675, and these 16675 events establish our control sample. In the further analysis, we use different subsets of the control sample, that satisfy the different criteria for the performed tests, as explained by the text.

\section{Results}\label{results}
After the visual control of both AGN and SNe is finalized, we have a final sample that contains the AGN-SN pairs that have passed through the filtering, see Table \ref{SupernovaCollection}. The selection process and the stages where visual control enters are shown in Figure \ref{Flowchart}. In total, we have 16 Type 1 AGN - SN pairs. In addition, Table \ref{Contaminants} shows the cases where there is uncertainty in whether the AGN is truly a Type 1 AGN, in SN type, redshift or if the SN is unconfirmed.


\begin{table*}
\caption{The main sample. The table shows all SN events that have been found in Type 1 AGN. Names, morphologies, host galaxy redshifts $z$, SN discovery references and types are listed. Also listed is the angular distance from the SN to the galaxy centre.}
\centering
{
\setlength\tabcolsep{2pt}
\begin{tabular}{|c c c c c c c c}
\hline\hline
\multicolumn{8}{c}{Collected sample} \\
\hline\hline
Galaxy name / ID & Activity class & z & SN name &  SN discovery & SN type & SN z  & d (arcmin)\\
NGC 5683 & Sy1 & 0.03664 & 2002db & \cite{Beutler2002} & Ia & 0.036218 & 0.2176552 \\  
NGC 3227 & Sy1 & 0.00365 & 1983U & \cite{Massone1983} & Ia & 0.0038 & 0.2805004 \\  
NGC 4619 & LINER (broadline) & 0.0233 & 2006ac & \cite{Lee2006} & Ia & 0.023106 & 0.3613452 \\  
UGC 8829 & LINERs (broadline) & 0.02612 & 2011O & \cite{Narla2011} & II & 0.0262 & 0.1389504 \\  
Anon J142135+5231 & NLSy1 & 0.24917 & 2003fh & \cite{Ellis2003} & Ia & 0.2486 & 0.0222 \\
NGC 4051 & NLSy1 & 0.00216 & 2003ie & \cite{Benetti2003} & II & 0.002336 & 1.550967 \\
" & NLSy1 & " & 2010br & \cite{Chomiuk2010} & Ib/c & 0.0023 & 0.2873169 \\
" & NLSy1 & " & 1983I & \cite{Wheeler1987} & Ic & 0.0023 & 0.8311077 \\
2MASX J03063958+0003426 & NLSy1 & 0.107379 & 2006np & \cite{Bassett2006} & Ia & 0.1064 & 0.09770007 \\
3XLSS J020514.8-045640 & NLSy1 & 0.362131 & ESSENCEn263 & \cite{Narayan2016} & Ia & 0.3677 & 0.04237974 \\
2MASS J00332804-0019129 & No blue bump Sy1 & 0.10675 & 2006ho & \cite{Frieman2006} & II & 0.1083745 & 0.0063 \\
NGC 4639 & Sy1 & 0.00364 & 1990N & \cite{Thouvenot1990} & Ia & 0.003397 & 1.055354 \\
NGC 7469 & Sy1 & 0.01588 & 2008ec & \cite{Harutyunyan2008} & Ia & 0.016317 & 0.2371698 \\
" & " & " & 2000ft & \cite{Colina2002} & II & 0.016 & 0.04557676 \\
NGC 931 & Sy1 & 0.016338 & 2009lw & \cite{Li2009} & Ib/IIb & 0.0167 & 0.5141631 \\
SDSS J123118.84+082315.4 & Sy1 & 0.16542 & 2008cb & \cite{Kasliwal2008} & Ia & 0.16535 & 0.04762122 \\
\hline
\hline
\end{tabular}
}
\label{SupernovaCollection}
\end{table*}


We start by examining the host galaxy properties of our main sample. After that, we investigate the typical AGN properties of our sample, and finally, the SN events and occurrences themselves.


\begin{table*}
\caption{The contaminant sample. The table shows all SN events that have been found in apparent Type 1 AGNs, but that may be unconfirmed or have dubious AGN/SN spectra. Names, morphologies, host galaxy redshifts $z$, SN discovery references and types are listed. In the OSC some of these host galaxy redshifts are missing, so we complete them with redshifts from SDSS and NED.} Also listed is the angular distance from SN to the galaxy centre.
\centering
{
\setlength\tabcolsep{2pt}
\begin{tabular}{|c c c c c c}
\hline\hline
\multicolumn{6}{c}{Collected sample} \\
\hline\hline
Galaxy name / ID & z & SN name & SN type & SN z  & d (arcmin)\\
2dFGRS TGN242Z167 & 0.15604 & Gaia14adg & II & 0.15 & 0.150\\
IC 2637	& 0.02925 & SN1989G & I(a?) & 0.029 & 0.033\\
Mrk 785	 & 0.04919 & SN1996ce & N.A. & 0.049 & 0.101\\
2MASX J23342408-0053250 & 0.08916 & SN2007ra & Ia & 0.089163 & 0.020\\
NGC 7589 & 0.02979 & SN2011hv & Ia & 0.0298 & 0.219\\
IC 3528 & 0.04582 & SN2001Z & II & 0.0455 & 0.186\\
Anon. 154413-0153 & 0.924 & SN2001cc & N.A. & N.A. & N.A.\\
2SLAQ J232546.49+001749.6 & 0.51348	& SN1999fq & N.A. & N.A. & 0.017\\
(VV2010c) J093313.9+015858 & 0.03112 & (GBM2015) SDSS475-51965-626SN & II & N.A. & 0\\
2MASS J08393992+3539151	& 0.16024 & (GM2013)SDSS0864-52320-82SN & II & N.A. & 0.002\\
NGP9 F270-0377962 & 0.05801 & (GBM2015)SDSS2102-54115-72SN & II & 0.05801 & N.A.\\
NGC 5548 & 0.01651 & 1984Z & II & 0.01727 & 0.089\\
Mrk 590 & 0.02609 & 2018djd & Ia & 0.026385 & 0.0261\\
2MASX J22423925+0118061 & 0.10666 & 1990ad & II & N.A. & 0.089\\
\hline
\hline
\end{tabular}
}
\label{Contaminants}
\end{table*}


\subsection{Host galaxy properties}

\subsubsection{Redshift distribution}

Figure \ref{fig:redshift} shows a histogram of the redshifts in our AGN-SN sample (the ``main'' sample) and a control sample of the OSC (the ''control'' sample) that has well identified host galaxies and with well-measured redshifts (only about $\sim$ 12000 events). We perform a two-sample KS test (using a nominal value $\alpha$ < 0.05 for ``significance'') on the two given distributions within this redshift range and see that the two distributions have no significant difference, where the p-value $\sim$ 0.53 is well above the nominal value. The main sample may be too small to detect any significant differences.

Therefore, we also have a look at what happens if we add the possible other Type 1 AGN, the objects that may either be real Type 1 AGN/SN pairs, or just contaminants. We plot the redshift distributions in Figure \ref{fig:redshiftcont}. Two-sample KS tests between the control sample and the larger ``mixed'' (main + contaminant) sample cannot defy the null hypothesis ($p \sim 0.45$), that the two samples are similar. 

Comparing the two samples, indicates there might exist some difference between the redshift distributions of the ``main sample'' and of the contaminants, as the Kolmogoroff-Smirnoff test yields $p \sim 0.062$, very close to the nominal value, but not significant.

\subsubsection{Morphologies of the host galaxies}

Using the morphologies listed in Galaxy Zoo and SIMBAD (see Section \ref{sec:control}) for both the control and main samples, we can compare the morphologies. We list them in Table \ref{Morphologies}. In the control sample, 24\% of the known host galaxies have information about their morphologies. In comparison, about half of the objects in the main sample do have the same information.
However, the difference is not statistically significant.

For those objects that have the host galaxies resolved in the control sample (3425 objects), we can compare the occurrence rates of different morphologies with the main sample. We see that ca 8\% reside in elliptical hosts in the control sample. In the main sample, the corresponding fraction of objects in elliptical hosts is 0\% (not a single elliptical host). A SN detection may bias the host galaxies towards being more star-forming, which can influence our statistics in favour of spiral hosts among the AGN/SN pairs. This can also be seen by the larger fraction of spiral hosts in the main sample ($\sim$ 46\%) versus control sample ($\sim$ 27\%).

\subsection{SN events}

\subsubsection{Crude estimates of detection rates}\label{sec:detections}

The star-formation history in a complete sample of galaxies can be seen by the absolute detection rates of SNe and the relative ratios of thermonuclear versus core-collapse SNe. Progenitors of core-collapse SNe are massive stars with short life times $< 10^{7}$ years\footnote{For single stars, the life times are $< 4 * 10^{7}$ years, while for binary stars they may last up to $3 * 10^{8}$ years. On a cosmic scale, marriage might not be that bad.}. These are indicative of recent or ongoing star-formation as in star-forming spiral galaxies \citep{vanBergh1991}. The core-collapse SN include classes like SNIb, SNIc (that show no Balmer lines), SNIIb (that show quickly fading Balmer lines), and SNIIn, SNII-P and SNII-L. Thermonuclear SNe, on the other hand, are indicators of past star formation in the earlier epochs, as their progenitors are white dwarfs that take about $10^{8} - 10^{10}$ years to form, with a peak around 1 Gyr. These have prominent silicate lines and characteristically declining light curves, and constitute one single big group: the SNIa.

In \cite{Villarroel2017}, we used SNe from the coherently collected intermediate Palomar Transient Factory (iPTF) to look at the stellar population in Type 1 AGN. However, they only found one single SN, which no statistics could be based on. We therefore, have a look at the detection rates of thermonuclear versus core-collapse SNe in our main sample. However, the main sample based on the OSC is not complete, even though it still may be indicative of certain trends. Table \ref{Classes} shows the distribution of different type of SNe in our main sample. We compute the fractions, effect sizes (see Table 3 and Section 5.2 \citep{Villarroel2017} for statistical treatment) and compare the ratios of thermonuclear SNe and core-collapse SNe for our main sample and our control sample. There appear to be some tendencies towards a higher occurrence number of SNII in the contaminant sample (and a higher number of SNIa in the main), although no statistically significant difference can be seen. Larger sample sizes are needed.

There have also been suggestions that Seyfert galaxies may have a higher fraction SN Ib and SN Ic over SN II, which indicate bursts of young star formation in Seyferts with average ages $\tau < 20$ Myrs. \cite{Bressan2002} report that normal spiral galaxies show a ratio $D_{Ib-c / II} \sim 0.23$, while the Seyfert galaxies have a significantly higher ratio. We therefore look at the numbers of these classes of SNe in Table \ref{Classes}. We find that the control sample has 639 SN Ib-Ic SNe and 3538 SN II SNe ($D_{Ib-c / II} \sim 0.181 \pm$ 0.008) The main sample has 3 SN Ib-Ic SNe and 4 SN II SNe ($D_{Ib-c / II} \sim 0.750 \pm 0.573)$. While we indeed see this trend, we cannot confirm it on a statistically significant level.

To compare the overall detection fractions, we make a back-of-the-envelope estimate of the number of galaxies that may host a SN bright enough to be observed by the typical SN surveys. Assuming the typical depth of a wide field survey like the ZTF\footnote{A wide field survey using a larger telescope, may of course find many supernovae at higher redshift An example is the Dark Energy Survey that has discovered over 2000 SNe at $z < 1$.} where the survey has a depth of 20 mag and weaker transients are not detected, most of the SNe such a survey finds have redshifts $z < 0.2$. We can now calculate the detection fraction in two different ways. In the first way, we calculate the co-moving volume for $z = 0.2$ which corresponds to about $2.1 \ast 10^{9}$ Mpc$^{3}$ using   Standard Cosmology. The galaxy density in this local region is about 0.01 galaxies Mpc$^{-3}$ \citep{Hill2010}. From this, the number of galaxies is about $10^{7}$ within this volume. About 0.70 $\ast$ 55000 SNe (the 70\% below redshift  $z = 0.2$ and $10^{7}$ galaxies, make a detection fraction of about $4 \ast 10^{-3}$ SNe per galaxy on average in the OSC. We find between 16 to 30 SNe on about 500 000 Type 1 AGN host galaxies the approximate total number from both the SDSS and SIMBAD parent samples, which at most corresponds to a detection fraction on about $6 \ast 10^{-5}$, about a hundred times less. Given that it is approximately two orders of magnitude lower than the expected average in the OSC, there seems to be a scarcity of reported SNe from Type 1 AGN host galaxies.

The second way to estimate the detection fraction, is with help of catalogued luminosity functions. We use the luminosity functions to calculate the number of Type 1 AGN and the total number of galaxies within a certain luminosity range at low redshift. The limiting magnitude of typical transient surveys cause the majority of SNe detected to have $z < 0.2$. We therefore investigate the numbers of host galaxies at this low redshift.

Using the luminosity functions for Type 1 AGN and ``all galaxies'' in the blue bands from Table 2 in \cite{Schulze2009} and Table 6 in \cite{Cool2012}, we estimate the expected ratio between Type 1 AGN and ``all galaxies'' at $z < 0.2$. We can do this by summing the number density per Mpc$^3$ per unit magnitude for each magnitude bin in a given magnitude interval. This summation is easy to do directly if the counts are complete over all luminosities. However, \cite{Schulze2009} observed an abrupt break in the optical/UV luminosity function at magnitudes $> -$19 that they believe is due to severe incompleteness of the HES sample at low luminosities. Moreover, the absolute magnitudes reported in \cite{Cool2012} relate to the magnitudes reported in \cite{Schulze2009} so that $M_{B,Cool} = M_{B,Schulze} - 5$ log $h$, which corresponds to a difference in 0.77 magnitudes for $h$ = 0.7 and also means that the Table 6 in \cite{Cool2012} actually starts at $M_{B} = -$18.77. At $M_{B} = -$19.5, the ratio between galaxies and Type 1 AGN is $\sim$ 400. At $-21.5 < M_{B} < -20$, the ratio is $\sim$ 700. At $M_{B}=-$ 22 the ratio approaches 1800, and at even higher luminosities, the ratio falls down again.

Therefore, we do the summation of number densities within the magnitude interval between $-22.5 < M_B < -19.50$. In this magnitude interval, we estimate the ratio of Type 1 AGN to ``all galaxies'' to be about one Type 1 AGN on 550 galaxies. Assuming that Type 1 AGN have the same SN detection rates as all galaxies, and that the true number of Type 1 AGN/SN events lies in between the numbers given by the main sample (16) and the mixed sample (30; main + contaminant), that means the expected number of SNe in the OSC should be anywhere in between 16 $\ast$ 550  to 30 $\ast$ 550 SNe, which translates to a sample between 8800 to 16500 SNe. This is a factor of three less than what actually has been reported in the OSC. This agrees with the conclusion from the first method of assessing the detection fractions where Type 1 AGN appear to show a deficit of SNe. This also agrees with the finding from \cite{Villarroel2017}, where only one SN was found in a sample of 10 000+ Type 1 AGN.

While a deficit of SNe near Type 1 AGN seems likely, none of the methods used is robust in estimating the total number of galaxies surveyed by the OSC.

\subsubsection{The location of SN events}
\label{sec:locations}
By looking at the radial distribution of the SN events inside the galaxies, and by looking at the inclination angle of the galaxy, we can learn whether the SNe explode at the same locations in Type 1 AGN as in galaxies ``on average''. Certain differences may be expected: on larger radii in the highly star-forming regions of the spiral arms, we expect more of core-collapse SNe. Also, if there is a strong bias that makes us miss most of the SN detections inside Type 1 AGN (as was suggested by the previous section \ref{sec:detections}), we expect to see a significantly bigger number of SN detections in the outer parts of the Type 1 AGN hosts, as compared to the control sample. 

We have a look at the inclination-corrected ratio $\phi_{0}$ between the distance of the SN from the center of the host galaxy $r$ and the angular radius R of the host galaxy, where
$\phi_{0}$ = $r / R$. The $R$ are given by the effective radius based on S\'{e}rsic profiles in the SDSS.

\begin{figure*}
	\includegraphics[scale=0.6]{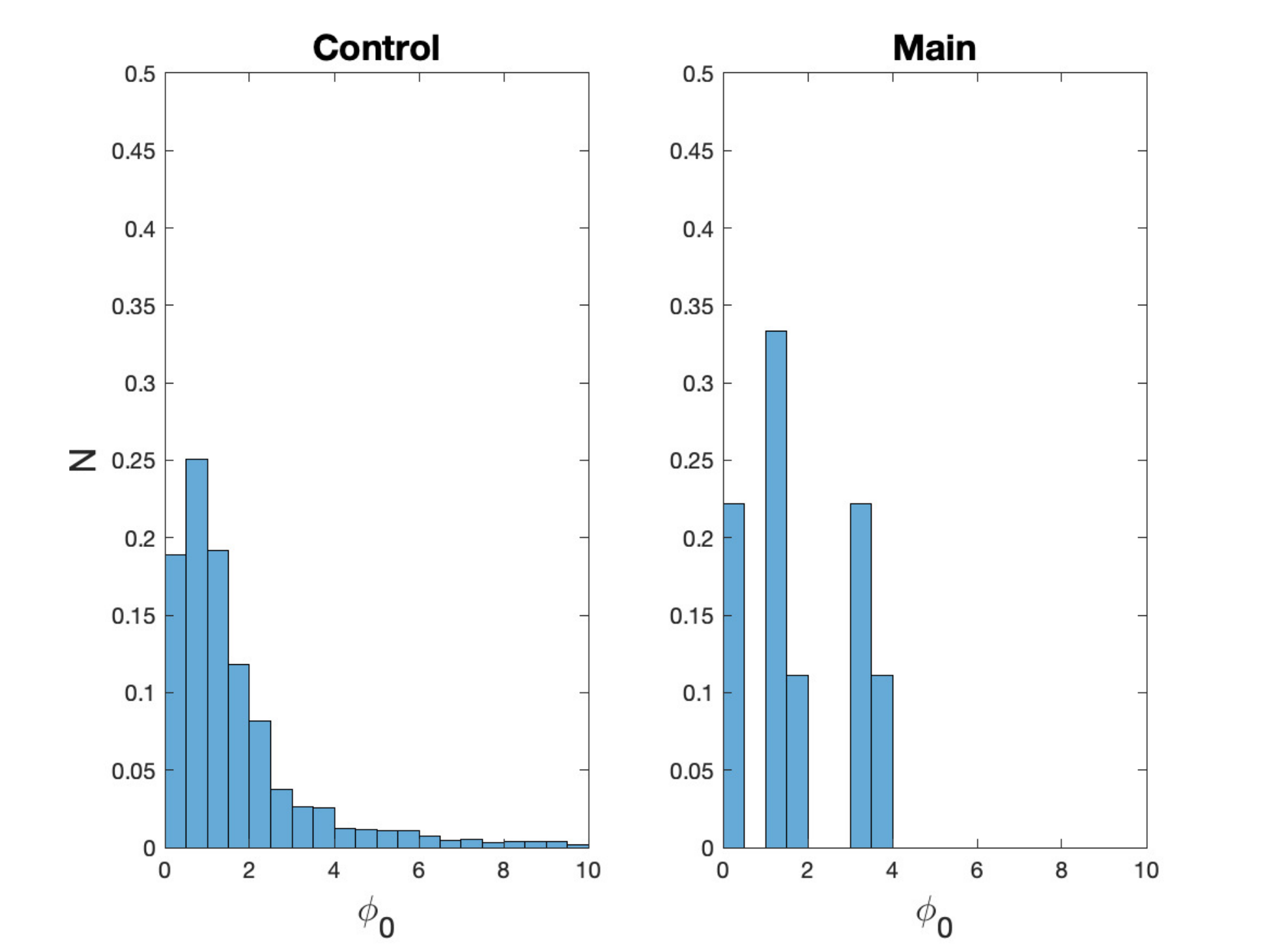}
    \caption{The ratio $\phi_{0}$ shows the approximate position of the SN inside the galaxy. Each plot uses the same bin width and normalisation ('N' is normalised counts). The left plot shows $\phi_{0}$ for all events in host galaxies with $\phi_{0} < 10$ in the control sample, about $\sim$ 2300 events. The right plot shows $\phi_{0}$ for the objects in the main sample with measured $\phi_{0}$. }
    \label{phiplot}
\end{figure*}

Figure \ref{phiplot} shows the $\phi_{0}$ in Type 1 AGN hosts and in the control sample, restricting ourselves to objects with $\phi_{0}$ $< 10$. A bright AGN may increase the $\phi_{0}$, and at a first glance it might look as if the main sample has slightly larger $\phi_{0}$, as compared to the control sample. But as we can see, the sample sizes of the Type 1 AGN hosts are too small to be able to detect any difference between
the samples, supported by the two-sampled Kolmogoroff-Smirnoff test ($p = 0.67$).

\section{Final remarks and Conclusions}

In a previous work \citep{Villarroel2017} we compared iPTF SNe in Type 1 and Type 2 AGN host galaxies and found a deficit of SN events in Type 1 AGN hosts (only one single SN was found). We have now searched for (nearly) all possible SN around Type 1 AGN in order to understand what caused the apparent deficit in the previous work. We find in total 30 Type 1 AGN/SN events from sampling 500 000 Type 1 AGN, which means we deal with small sample sizes. We examine the AGN/SN events. The conclusions are as follows:

\begin{enumerate}
    \item We see no statistically significant difference between the redshift distributions of Type 1 AGN/SN pairs and the control sample (representing the OSC).
    \item We see no statistically significant differences in the morphologies observed between the Type 1 AGN/SN pairs and the control sample that represents the typical SN hosts in the OSC.
    \item see no statistically significant difference in the location of SN events inside each galaxy, as seen by the inclination-corrected ratio $\phi_{0}$.
    \item There may be some differences in morphologies with the ``typical'' Type 1 AGN found in the Galaxy Zoo, see e.g. \cite{Villarroel2017,Chen2017}. Possibly, the prior requirement of a SN detection biases the morphology distribution towards more star-forming, late-type galaxies.

\end{enumerate}

In \cite{Villarroel2017} we observe that SNe are less common in Type 1 hosts
as we find only one SNe in our Type 1 AGN sample. The estimated difference is a
a factor of $\sim$ 10 compared to Type 2 AGN hosts or normal galaxies. We have now collected a sample of 500 000 optically selected Type 1 galaxies and broad line quasars. The detection fraction can be calculated with certainty for these Type 1 AGN by simply taking the ratio of the number of SNe we have and dividing by the number of AGN we have surveyed, see Section \ref{sec:detections}. But doing a similar exercise for the OSC sample yields large uncertainties when comparing to the entire OSC. We may have a fixed number of host galaxies collected in the OSC, but we do not have
a well-determined estimate of the number of galaxies the OSC actually includes. We try two different ways of estimating the number of galaxies. By carefully avoiding the lowest luminosities where the Type 1 AGN luminosity function in \cite{Schulze2009} suffers from incompleteness (as noted by the authors), we find that both methods imply that fewer SNe are detected near Type 1 AGN than in other galaxies in the OSC, supporting the observation of a deficit in \cite{Villarroel2017}. Considering the uncertainty in the detection fraction calculations, this inference should be viewed as amply preliminary.

From points (i) - (iv) we can state there is no general bias against the detection of SNe in a ``typical'' Type 1 AGN. But could there be a bias against detecting supernovae in Type 1 AGN when the Type 1 AGN also show strong variability? Or could supernovae and AGN variability be mixed up? Such objects could be mistaken for SNe in Type 1 AGN. For instance, be so-called ``changing-look AGN'' that switch between classes. Examples of changing-look AGN are those most extreme, where the broad-lines either appear or disappear entirely influencing the objects Type 1/Type 2 classification -- the SN hosts Mrk 590 and NGC 1018 among them, accompanied by changes in flux up to a factor of hundreds (as compared to just a fractional change in flux as commonly in ``normal'' AGN). But many other changing-look AGN jump between various intermediate classes.

We tried to examine whether there is a connection between the ``cleanliness'' of the Type 1 AGN classification and the SN classes. The Main sample are those Type 1 AGN that have secure classifications, while the Contaminants are those that probably are Type 1 AGN. We note, that the latter group have a larger incidence of core-collapse supernovae, see Tables \ref{SupernovaCollection} and \ref{Contaminants}. That could mean that core-collapse supernovae going off in the nuclear region of normal galaxies or Type 2 AGN sometimes may give rise to a spectrum similar to Type 1 AGN due to line broadening effects. An example where this might be happening is the case of the two dubious Type-1 AGN in Section \ref{comments} that can be explained as normal galaxies or Type 2 AGN contamined by core-collapse supernovae from the nuclear regions. Or, simply as observed in other works, Type 2 AGN and normal galaxies may have a higher incidence of core-collapse SNe. However, the effect is statistically insignificant with our small samples.

 We find a similar ratio of thermonuclear vs core-collapse supernovae in Type 1 AGN and in the control sample. Conclusively, the Type 1 AGN-SN sample we have constructed is consistent with the control sample in the properties of its host galaxies, and is a suitable for different follow-up studies that need both the SN and the AGN.


\begin{table*}[t]
	\centering
	\caption{Morphologies. We here compare the different morphologies of the 13 host galaxies represented in our main sample, our mixed sample (main + contaminants) and the control sample. A big fraction of the galaxies in the control sample have unknown morphology, and a small part among them (ca 81) have irregular morphologies. The total number of objects in each sample is give in in the parenthesis of the header, and the fractions of galaxies of a certain morphology are calculated with respect to the total numbers. The errors of the fractions are calculated using Poisson statistics.}
	\label{Morphologies}
	{
\setlength\tabcolsep{2pt}
\begin{tabular}{c c c c c c c} 
		Morphology & Main (13) & $f$ & Mixed (27) & $f$ & Control (14145) & $f$\\
		\hline
		Spirals (incl. S0) & 7 & 0.54 $\pm$ 0.25 & 12 & 0.44 $\pm$ 0.15 & 3160 & 0.223 $\pm$  0.004\\
		Elliptical  & 0 & 0 & 1 & 0.04 $\pm$ 0.04 & 265 & 0.019 $\pm$ 0.001\\
		Irregular or Unknown  & 6 & 0.46 $\pm$ 0.23 & 14 & 0.52 $\pm$ 0.17 & 10720 & 0.758 $\pm$ 0.001\\
		\hline
	\end{tabular}
	}
\end{table*}

\newpage
\begin{table*}
	\centering
	\caption{SN classes. The table shows the SN types from the main sample, the mixed (main + contaminants) and the control sample for events with known host galaxies. We have a look at all the SNe in the control sample. When we compare the fraction of thermonuclear vs core-collapsed SNe, we compare only look at the fractions of the classified (known) SN types. For the mixed sample, we thus omit the ``unknown'' and the 'I' supernova we cannot say if thermonuclear or core-collapse.}
	\label{Classes}
	{
\setlength\tabcolsep{2pt}
\begin{tabular}{c c c c c c c} 
		\hline
		SN type & Main (16) & $f$ & Mixed (30) & $f$ & Control (16675) & $f$\\
		\hline\hline
		I & 0 & 0 & 1 & 0.03 $\pm$ 0.03 & 83 & 0.005 $\pm 5*10^{-4}$\\
		Ia & 9 & 0.563 $\pm$ 0.23 & 12 & 0.400 $\pm$ 0.14 & 5550 & 0.333 $\pm$ 0.005 \\
		Ib & 0 & 0 & 0 & 0 & 179 & 0.011 $\pm 8*10^{-4}$\\
		Ib/c & 1 & 0.063 $\pm$ 0.06 & 1 & 0.03 $\pm$ 0.03 & 155 & 0.009 $\pm 7.5*10^{-4}$\\
		Ibn & 0 & 0 & 0 & 0 & 13 & $<$0.001 $\pm 2*10^{-4}$ \\
		Ib/IIb & 1 & 0.063 $\pm$ 0.06 & 1 & 0.03 $\pm$ 0.03 & 4 & $<$0.001 $\pm 1.2*10^{-4}$ \\
		Ic & 1 & 0.063 $\pm$ 0.06 & 1 & 0.03 $\pm$ 0.03 & 301 & 0.018 $\pm$ 0.001\\
		II & 4 & 0.25 $\pm$ 0.14 & 11 & 0.37 $\pm$ 0.13 & 3538 & 0.212 $\pm$ 0.004\\
		IIP & 0 & 0 & 0 & 0 & 266 & 0.016 $\pm 9.9*10^{-4}$ \\
		IIL & 0 & 0 & 0 & 0 & 3 & $<$0.001 $\pm 1*10^{-4}$ \\
		IIb & 0 & 0 & 0 & 0 & 91 & 0.006 $\pm 5.7*10^{-4}$ \\
		IIn & 0 & 0 & 0 & 0 & 141 & 0.009 $\pm 7.1*10^{-4}$ \\
		Superluminous & 0 & 0 & 0 & 0 & 17 &  0.001 $\pm 2.5*10^{-4}$ \\
		Unknown & 0 & 0 & 3 & 0.100 $\pm$ 0.06 & 5759 &  0.345 $\pm$ 0.005\\
		\hline
		Total thermonuclear & 9 & 0.46 $\pm$ 0.16 & 12 & 0.44 $\pm$ 0.15 & 5550 & 0.542 $\pm$ 0.009 \\
		Total core-collapse & 7 & 0.54 $\pm$ 0.18 & 14 & 0.56 $\pm$ 0.18 & 4691 & 0.458 $\pm$ 0.008\\
		\hline
	\end{tabular}
	}
\end{table*}

\newpage

\section*{Acknowledgements}

We thank the anonymous referee for the constructive comments. B.V. is funded by the Swedish Research Council (Vetenskapsr\aa det, grant no. 2017-06372). E.L. was partially supported by a European Union COFUND/Durham Junior Research Fellowship (under EU grant agreement no. 609412) and by an International Engagement Durham University Grant 2018. B.V. wishes to thank K. Oh for providing an AGN sample with carefully remeasured line widths. She also wishes to thank Anders Nyholm, Joel Johansson, Andreas Korn, Mart\'{i}n L\'{o}pez-Corredoira, Martin Sahlen, Martin Ward and C\'{e}sar Franck for fruitful discussions.

The CSS survey is funded by the National Aeronautics and Space
Administration under Grant No. NNG05GF22G issued through the Science
Mission Directorate Near-Earth Objects Observations Program.  The CRTS
survey is supported by the U.S.~National Science Foundation under
grants AST-0909182 and AST-1313422.

Funding for the Sloan Digital Sky Survey IV has been provided by the Alfred P. Sloan Foundation, the U.S. Department of Energy Office of Science, and the Participating Institutions. SDSS-IV acknowledges
support and resources from the Center for High-Performance Computing at
the University of Utah. The SDSS web site is www.sdss.org.

SDSS-IV is managed by the Astrophysical Research Consortium for the Participating Institutions of the SDSS Collaboration including the Brazilian Participation Group, the Carnegie Institution for Science, 
Carnegie Mellon University, the Chilean Participation Group, the French Participation Group, Harvard-Smithsonian Center for Astrophysics, Instituto de Astrof\'isica de Canarias, The Johns Hopkins University, Kavli Institute for the Physics and Mathematics of the Universe (IPMU) / University of Tokyo, the Korean Participation Group, Lawrence Berkeley National Laboratory, Leibniz Institut f\"ur Astrophysik Potsdam (AIP), Max-Planck-Institut f\"ur Astronomie (MPIA Heidelberg), Max-Planck-Institut f\"ur Astrophysik (MPA Garching), Max-Planck-Institut f\"ur Extraterrestrische Physik (MPE), 
National Astronomical Observatories of China, New Mexico State University, New York University, University of Notre Dame, Observat\'ario Nacional / MCTI, The Ohio State University, Pennsylvania State University, Shanghai Astronomical Observatory, United Kingdom Participation Group, Universidad Nacional Aut\'onoma de M\'exico, University of Arizona, University of Colorado Boulder, University of Oxford, University of Portsmouth, University of Utah, University of Virginia, University of Washington, University of Wisconsin, Vanderbilt University, and Yale University.









\bsp	
\label{lastpage}
\end{document}